\documentclass[preprint,aps,,showpacs,amsmath,amssymb]{revtex4}

\usepackage{graphicx}
\usepackage{dcolumn}
\usepackage{bm}

\begin{document}
\title{
 Theory of anomalous diffusive reaction rates on realistic self-affine fractals
}
\author{ Rama Kant}
\affiliation{Complex Systems Group, Department of Chemistry, \\
University of Delhi, \\
Delhi 110007, India}
\date{\today}

\begin{abstract}
{
In this letter, a theoretical method for the analysis of diffusive flux/current to limited scale self-affine 
random fractals is presented and compared with experimentally measured electrochemical current for such roughness. 
The theory explains the several experimental findings of the temporal scale invariance 
as well as deviation from this of 
current transients in terms of three dominant fractal parameters for the limited-length scales of roughness.  
This theoretical method is based on limited scale power-law characterization of the interfacial
roughness spectrum and the solution of diffusion equation under the diffusion-limited boundary conditions 
on rough interfaces.  More broadly, these calculations challenges the earlier belief
that the anomalous behavior is solely dependent on fractal dimension of roughness and highlight the potential
to apply this equation for the scale invariant roughness determination. Finally, the validity of theoretical result 
is tested with extensive experimental data.}

\end{abstract}

\maketitle

Realistic surface roughness has limited-length scales of irregularities and are frequently 
characterized as self-affine fractals \cite{feder,Mandelbrot84,Sayles78,Zaiser04}. Diffusion-limited 
processes on such interfaces show anomalous behavior of the reaction flux. Some of the diverse realizations 
of diffusion-limited processes in physical phenomena are: Spin relaxation\cite{de-gennes}, 
fluorescence quenching\cite{de-gennes,kopelman}, heterogeneous catalysis\cite{pfeifer84,Lee2002}, 
enzyme kinetics\cite{pfeifer85,Dewey}, heat diffusion\cite{Roux95:2}, membrane transport\cite{sapoval91,Sapoval96} 
and electrochemistry\cite{LeMehaute-Crepy83,Pajkossy94,Kant93,Kant94:2,Kant97,Pajkossy91,Halsey91:92,Ocon-Arvia91,Go-Pyun2005}. 

The diffusion of the reactant from a bulk medium towards an interface where the reactants either loose
their activities or are transformed into product is a common problem in diverse areas of 
science \cite{de-gennes,kopelman,Lee2002,pfeifer84,pfeifer85,Dewey,Roux95:2,sapoval91,Sapoval96,sapoval06,LeMehaute-Crepy83,
Pajkossy94,Halsey91:92,Kant93,Kant94:2,Kant97,Pajkossy91,Ocon-Arvia91,Go-Pyun2005,Ball02}.
These processes are experimentally approximated as power-law relation of reaction
rates/flux, $J(t)$, in time ($t$) and is represented by the following relation:
\begin{equation}
J(t)\sim t^{-\beta} \label{de-gennes-flux}
\end{equation}
where the exponent, $\beta$ depends on interfacial roughness. Theoretical justification for Eq.(\ref{de-gennes-flux})
was provided by De Gennes scaling result with $\beta =(D_H-1)/2$\cite{de-gennes} and generalized form\cite{Maritan}. DeGennes analysed it
for the problem of 
diffusion-controlled nuclear magnetic relaxation in  porous media  with fractal 
interfacial dimension $D_H$. Later, similar results were discussed for other diffusion-controlled situations, 
such as adsorption on a porous fractal catalyst\cite{pfeifer84}, 
in the context of flow of energy and mass through a fractal interface\cite{LeMehaute-Crepy83},
for rough fractal 
electrode/electrolyte interfacial current under potentiostatic conditions\cite{Pajkossy94} and similar result for
heat diffusion from a self-affine fractal boundary\cite{Roux95:2}. Due to its simplicity and the lack of better 
alternative of Eq.~(\ref{de-gennes-flux}) captured lots of attention and is used indiscriminately 
for interpreting large quantity of data \cite{Pajkossy91,Ocon-Arvia91}. 
These include cases where the range of roughness is too small to be taken as 
idealized fractals. It is commonly known that Eq.~(\ref{de-gennes-flux}) is 
unable to include complete characterization of realistic fractal roughness and also does not reproduce subtle aspect 
of experimental data.

I, present an analytical model that allows one to probe the anomalous time behavior of diffusive flux 
of reacting interface with roughness and to evaluate the consequences of limited length scales of roughness 
on the anomalous behavior. This theory explains the several experimental findings \cite{Pajkossy91,Ocon-Arvia91} 
of the temporal scale invariance of 
flux/current in terms of three dominant fractal parameters for the limited-length scales of roughness.

 The information about surface roughness enters in my theory  through power spectrum of roughness.
The power spectrum of a realistic surface (also called "approximate self-affine fractal")
is described in term of limited scales of wave-numbers ($K$) power-law function\cite{Sayles78,yordanov02} 
i.e., $\left<\left|\hat{\zeta}(\vec{K})\right|^2\right> =T\, |K|^{2D_H-7},\;{\rm for} \;  1/{\it L}\le |K| \le 1/\ell$.
There are four physical parameters of roughness in this framework, namely $D_H$, ${\ell}$ , $\it L$ and $T$.
$D_H$ is the fractal dimension,  a global property which describe scale invariance property of the roughness-
an anomalous behavior in flux and its time exponent is usually assumed to be function of this parameter; 
${\ell}$ and $L$ are lower and upper cutoff length scales of fractality, respectively; and $T$ is the strength of fractal 
and related to topothesy of fractals\cite{Sayles78,berry,yordanov02}, its units are ${\rm cm}^{2D_H-3}$ and $T\rightarrow 0$ implies no roughness.
The lower roughness scale is the length above which surface show fractal behavior.  Also, such power-law spectrum  may represents
band-limited form of Ausloos-Berman's generalization\cite{Ausloos-Berman} for multivariate Weierstrass-Mendelbrot function for an isotropic (statistically) rough surface. 
The moments of power-spectrum are related to various morphological features of rough surface {\em viz.} 
rms width($\sqrt{m_0}$), rms gradient($\sqrt{m_2}$), rms curvature($\sqrt{m_4}$) etc. 
The general moments of power 
spectrum  (i.e., 2k-th moments, $m_{2k}$) are easily obtained for above mentioned power-spectrum
and are important morphological characteristic of surface roughness. The general formula is:
$m_{2k} = {T}\left({{\ell^{-2\delta_k}} - {L^{-2\delta_k}}} \right)/{4\pi}{\delta_k }$,
where $\delta_k = \delta+k$ and $\delta =D_H-5/2$.

Present theory of anomalous diffusive reaction rates on realistic self-affine fractals is obtained using our 
 general formalism \cite{Kant93,Kant94:2} which 
show the diffusive flux/current at random rough surfaces can be described in term of its 
power spectral density. Main approximation involved in derivation of general formalism 
is the truncation of solution at second order in surface roughness profile\cite{Kant93,Kant94:2}.
The total (averaged) flux/current at the stationary, Gaussian random surface is given by\cite{Kant93,Kant94:2}:

\begin{equation}  
\hskip -0.2in J(t)   = \frac{D\; A_0\; C_s}{\sqrt{\pi\,D\, t}} \left[ 1 + \frac{1}{4\pi Dt} \int_0^\infty dK K\left( 
1 - e^{-K^2  Dt} \right) \left<\left|\hat{\zeta}(\vec{K})\right|^2\right> \right] \label{mcps}
\end{equation}
where $D$ is the bulk molecular diffusion constant, 
 $C_s$ is the difference between the surface and bulk concentration and $A_0$ is the area of surface around which rough surface fluctuates.
 The diffusion controlled reaction rates are related to potentiostatic current transients of an electrode
undergoing fast charge transfer. The reaction flux, $J(t)$, is related to electrode current ($I(t)$) as: 
$I(t)=-nFJ(t)$, where $n$ is the number of electron transfer in redox reaction 
($O_{Solution}+ne^- \leftrightharpoons R_{Solution}$) and $F$ is Faraday constant.

The exact solution for the dynamic diffusive flux on an approximate self-affine surface 
(substituting above mentioned band-limited power law spectrum in Eq.~(\ref{mcps}))
under diffusion-limited condition is obtained as
\begin{equation}
\hskip -0.4in J(t)= \frac{D\; A_0\; C_s}{\sqrt{\pi\,D\, t}}
  \left(1 + \frac{T}{8 \pi} \left(\frac{ {\ell}^{-2\delta}-{L}^{-2\delta}}{\delta Dt}
+\frac{\Gamma\left(\delta,\;Dt/\ell^2,\; Dt/L^2\right)}{(D t)^{1+\delta}}\right) \right) 
\label{exact}
\end{equation}
where $\delta =D_H-5/2$, $\Gamma \left(\alpha ,x_0, x_1\right)=
\Gamma \left(\alpha ,x_0\right)-\Gamma \left(\alpha ,x_1\right)=
\gamma \left(\alpha ,x_1\right)-\gamma \left(\alpha ,x_0\right) $,
$\Gamma (\alpha,x_i)$ and $\gamma (\alpha,x_i)$ are the incomplete Gamma 
functions \cite{abramo}.  
Equation~(\ref{exact}) graphically analysed in Fig.1 which show dependence of scaling region
on $\ell$, $D_H$ and $T$. The scaling region has very weak dependence on $L$ so it is not shown in Fig. 1. 

Most of experimental data recorded are for 
the intermediate time regime i.e., $L^2/D\, \gtrsim \, t\, \gtrsim \ell^2/D$.  
The expression for intermediate time is obtained using time constraints in 
expansion of two incomplete gamma functions and retaining only leading orders.
 Final equation has a simple and elegant form as follows:
\begin{eqnarray}
J(t)&\approx& \frac{D\; A_0\; C_s}{\sqrt{\pi\,D\, t}}
  \left(1 + \frac{T}{8 \pi} \left(\frac{ {\ell}^{-2\delta}}{\delta Dt}
-\frac{\Gamma\left(\delta\right)}{(D t)^{1+\delta}}\right) \right) 
\label{scale-II}
\end{eqnarray}
which capture the anomalous behavior of  reaction flux. $\Gamma(x)$ in Eq.(\ref{scale-II}) is the gamma function\cite{abramo}.
The small length scale of fractal roughness command the flux transient for roughness 
of limited scales, i.e., ${\it L} /{\ell} \sim 10$. There is no dependence on the upper length scale(${\it L}$) for fractality 
in Eq.~(\ref{scale-II}).

Equations (\ref{exact}) and (\ref{scale-II}) extends the conventional representation of the Cottrell current 
transient ($~1/\sqrt{t}$) on the planar electrode in electrochemistry\cite{Bard} 
to the fractally rough electrode. This equation achieves a more realistic characterization 
of limited scale rough surface diffusive flux as it includes the fractal dimension dependent 
power-law as well as the contribution from the length scale and strength of fractality. 
The total mean flux is the summation of smooth surface 
flux and an anomalous excess flux due to fractal roughness or can also be looked upon as product of the $1/\sqrt{t}$ 
current and dynamic roughness factor (term inside parenthesis of Eqs.~(\ref{exact}) and (\ref{scale-II})).

\begin{figure}[htbp]
\includegraphics[width=6cm,height=8.5cm]{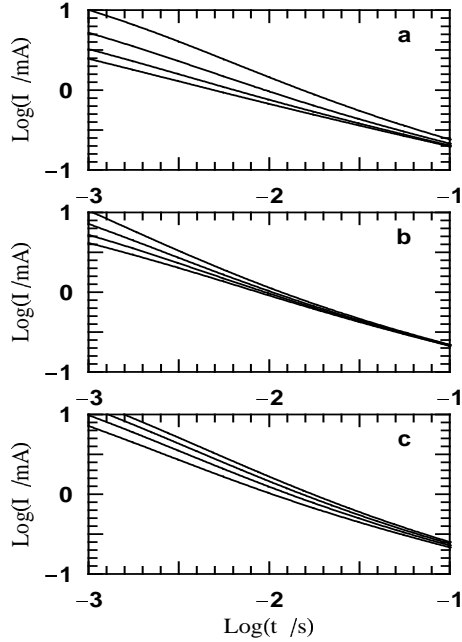}
\caption{\label{figT1} 
{Illustrates the effect of three dominant fractal parameters, i.e., $D_H$, ${\ell}$ and $T$, on the
anomalous scaling behavior} of reaction rates. These theoretical results are plotted as graphs to 
show an approximate power-law dependence of the 
flux $\slash$ current on  time. These calculations break the earlier beliefs based on idealized fractal models 
that the exponent of anomalous diffusion region purely depends on fractal dimension of roughness. 
 {\bf a}, {Anomalous region dependence on the fractal dimension of roughness.} 
The solid line is generated using ${\ell}=0.6 \mu m$, $T=1.3 \times 10^{-6}(arb. \; units)$ and $D_H$ = 2.45, 2.40, 
2.35 and 2.30 from above. Ratio of ${\it L }/{\ell }$ is kept fixed (=100) in all calculations.
{\bf b}, { The effect of lower scale of roughness on the anomalous region} is clearly depicted in the double logarithmic  
plot of current transients. 
The solid line is generated using $D_H=2.4$, $T =1.3 \times 10^{-6}(arb. \; units)$ and ${\ell}$ = 0.2$\,\mu m$, 0.4$\,\mu m$, 
0.6$\,\mu m$ and 0.8$\,\mu m$ from above.
{\bf c}, { Shows the effect of strength of fractal on the anomalous region} or slope of double logarithmic plot of current transients. 
The solid line is generated using $D_H=2.4$, ${\ell}$ = 0.4$\,\mu m$ and $T=T_0=1.3 \times 10^{-6}\,(arb. \; units)$, $1.5\times T_0$,
$2\times T_0$ and $2.5\times T_0$ from above. 
Other fixed quantities in calculations are: macroscopic areas ($A_0=0.1\; cm^2$), 
diffusion coefficient ($D=5\times 10^{-6} \;cm^2/s$) and concentration ($C_O=C_R=5\; mM$). 
These figures clearly demonstrate that the exponent of the anomalous region is dependent on  all three fractal roughness parameters.
}
\end{figure}

Nature of the plots also elucidates anomalous scaling behavior in the intermediate time regime.
The current transients increased with decrease in small scale of roughness in the early time domain, 
which is similar but larger in the magnitude of slope as compared to the planar $1/\sqrt{t}$ response.
It follows a power law behavior in the intermediate time regime which merges with large time $1/\sqrt{t}$ behavior.
As the range of roughness increases, the roughness factor and total current output also 
increases simultaneously on small time scales. So one can say that the total current output at small 
times is dependent upon the lower cutoff length scale ${\ell}$. No such impact of ${\ell}$ 
has been observed at large time scale as this regime is controlled by the width of interface $m_0$.
The width of interface is a strong function of ${\it L}$ and a weak function of ${\ell}$. 
Another important feature that can be verified from 
these graphs is that there is no such sharp outer cut-off time but the inner cut-off time decreases 
with the decrease in the lower length scale of roughness or with the increase in roughness factor.

What is remarkable about equation~(\ref{scale-II}) is that it governs the diffusive 
flux of the diverse set of roughness features. Its validity is not 
only tested, in this letter, for various range of roughness but also for the magnitude of roughness factor too. 
The conventional scaling equation (1) cannot explain deviation from
linear behavior in log-log plot which is often seen in data\cite{Pajkossy91,Ocon-Arvia91}
and is easily captured by this theory. 
Figure (2)  has several current transient curves and experimental data\cite{Ocon-Arvia91} 
on rough electrodes with limited length scales of fractality. 
Data are analyzed keeping in mind the experimental information about roughness factor that is available about their roughness.
Two lengths ${\ell}$ and ${\it L}$ are approximately identified from STM image of surface\cite{Ocon-Arvia91}. 
Identification of $D_H$, $\ell$ and $T$ is achieved by the minimization of deviation from experimental time-dependent data of 
diffusion-limited current.

\begin{figure}[htbp]
\includegraphics[width=6cm,height=4cm]{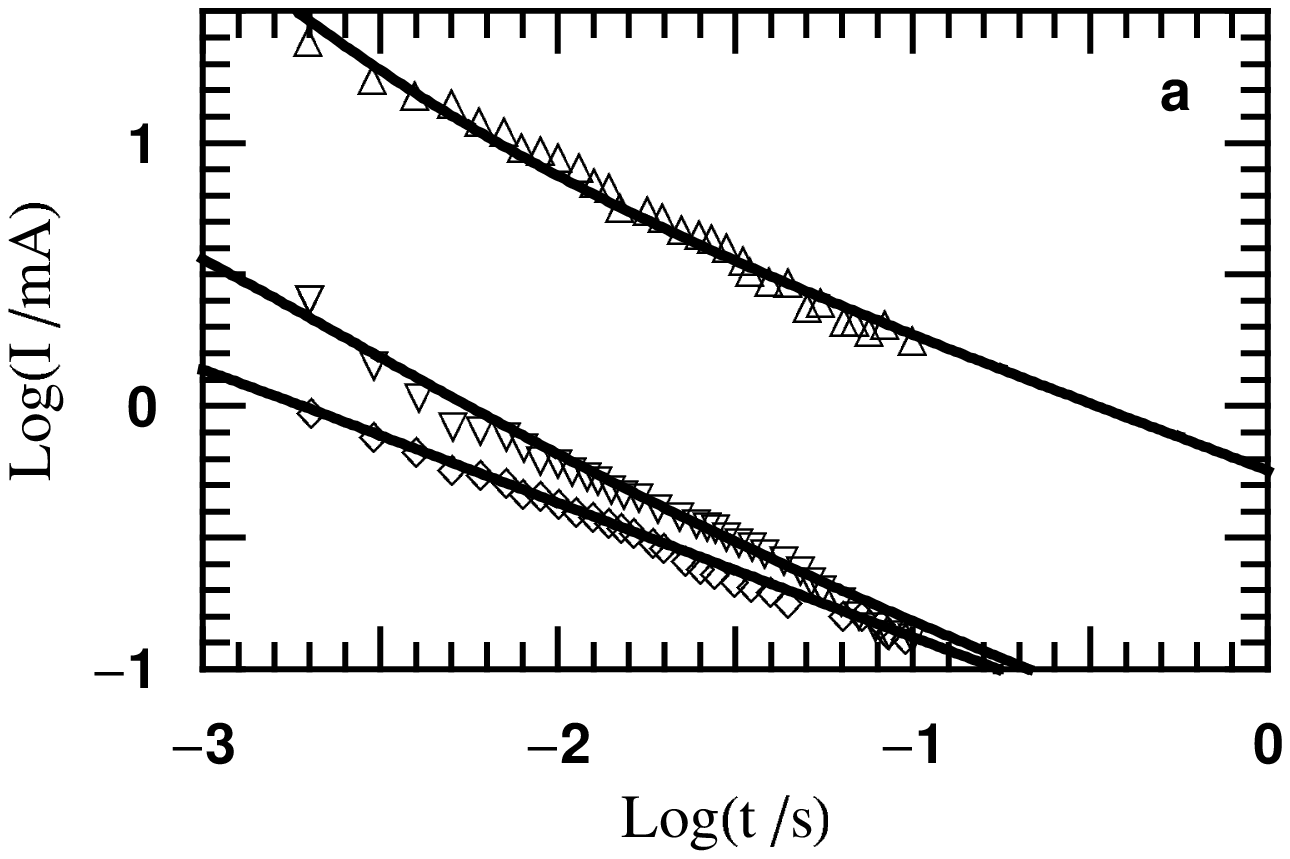}
\includegraphics[width=6cm,height=4cm]{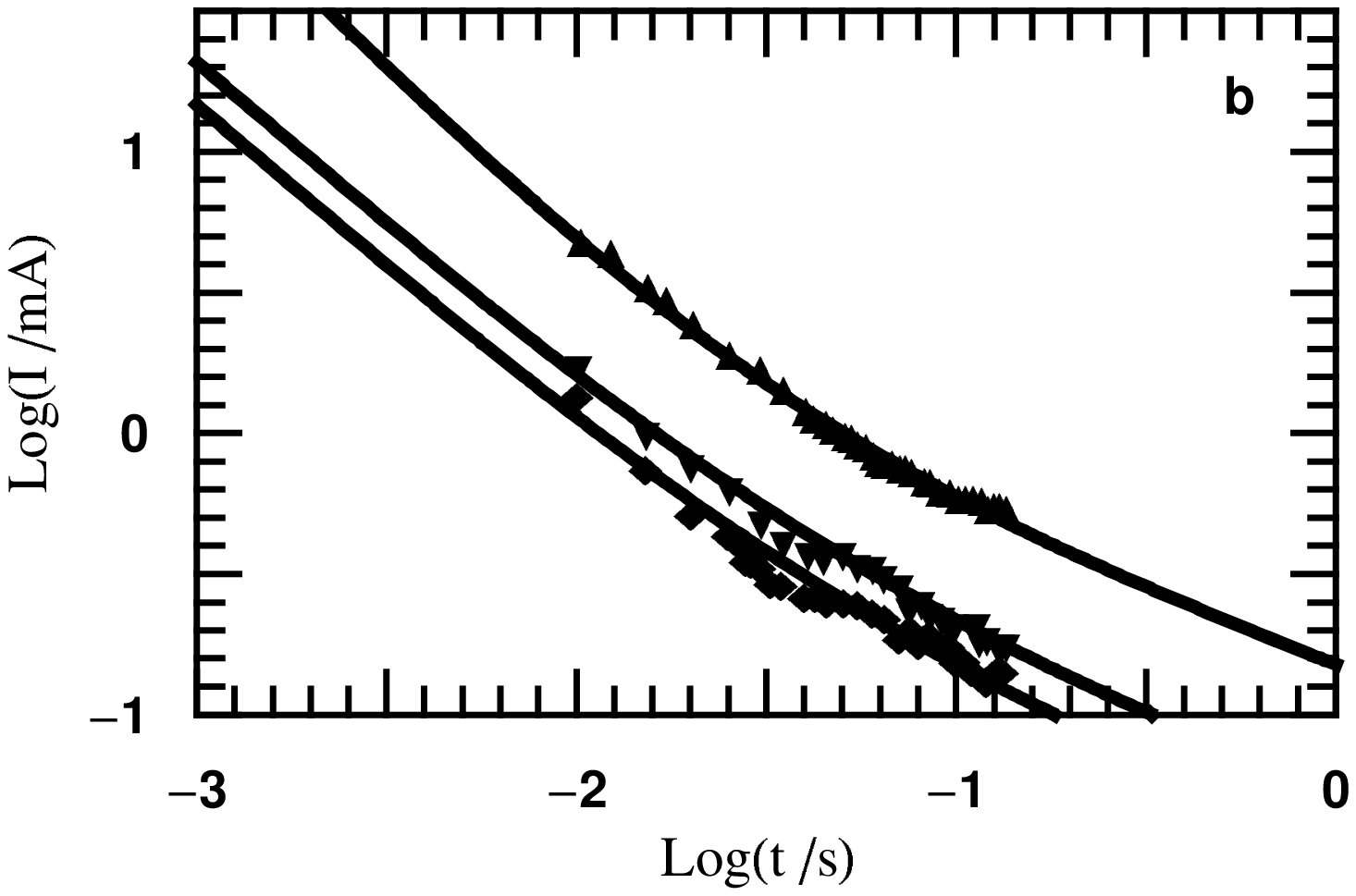}
\caption{\label{figE1} 
{ Comparison of model predictions from Eq.~(\ref{scale-II}) and current transients
data of Ocon et al\cite{Ocon-Arvia91}.} 
Electrodes in these diffusion-limited current experiments used were nano scale electrodispersed 
gold wire with surface layer of thin columnar gold\cite{Ocon-Arvia91}.
There roughness is changed by ageing of these columnar structure gold deposits by a slow reorganization 
processes. Ageing may lead to increase in $\ell$ and decrease in $D_H$ of rough surface. 
Rough surface of a wire electrode is imagined to be randomly fluctuating 
surface around a macroscopic plane as the curvature contribution of macroscopic wire geometry is insignificant.
{\bf a}, The solid lines are generated using: (1) $D_H=2.42$, $\ell=11\; nm$,  $T=2.0 \times 10^{-7} (arb. \; units)$, $A_0=0.93\; cm^2$,  
and experimental data points ($\triangle$).
(2) $D_H=2.19$, $\ell =400\; nm$,  $T=4.0 \times 10^{-5}(arb. \; units)$, $A_0=0.073\; cm^2$ 
and experimental data points ($\triangledown$).
(3) $D_H=2.02$, ${\ell}=750\; nm $, $T=5.0 \times 10^{-5} (arb. \; units)$, $A_0=0.068\; cm^2$  and experimental data points ($\lozenge$).
{\bf b}, The solid line is generated using: (1) $D_H=2.52$, ${\ell}=20\; nm$,  $T=1.94 \times 10^{-7} (arb. \; units)$, $A_0=0.25\; cm^2$ 
and experimental data ($\blacktriangle$).
(2) $D_H=2.33$, ${\ell} =86\; nm$,   $T=1.04 \times 10^{-5} (arb. \; units)$, $A_0=0.085\; cm^2$ and experimental data ($\blacktriangledown$).
(3) $D_H=2.32$, ${\ell} =87\; nm$,  $T=1.32 \times 10^{-5} (arb. \; units)$, $A_0=0.060\; cm^2$ and experimental points ($\blacklozenge$).
}
\end{figure}

Surprisingly, limited order perturbation analysis is able to capture the features of large roughness too.
Though one expects that the scaling results would be seen  only in large roughness form of theory to 
match with assumptions of Eq.~(\ref{de-gennes-flux})\cite{Kant97}. Most important of all, this work shows 
an intermediate anomalous power-law form for time above inner transition time ($t_i$).
This suggests that this theory have extended range of validity much beyond  expectations 
which is also seen in comparison with experimental results in Fig.~(\ref{figE1}). 
Another feature of this theory is that $t_i$ decrease with decrease in ${\ell}$. Inner transition time is evaluated equating small time expansion 
\begin{equation}
J(t)=\frac{D\; A_0\; C_s}{\sqrt{\pi\,D\, t}} \left( 1 + \frac{m_2}{2} - \frac{m_4}{4} Dt  
+  \cdots \right)\label{short}
\end{equation}
and Eq.~(\ref{scale-II}) at  $t=t_i$. 

Results developed in Eqs.(\ref{exact}) and (\ref{scale-II}) are based on statistical models for the current on random surface fractals. 
Advantage of such formulation is that it is based on 
four statistical parameters: $D_H$, $T$, ${\ell}$ and ${\it L}$. Most cases, we do not have independent information about fractal parameters as experimental studies 
rarely measure roughness power spectrum. In some cases, small and large length scales of roughness is characterised from
SEM or STM image of surface. Similarly, the knowledge of surface roughness factor or width of roughness or both can
help to fix these statistical parameters.
The remaining unknown parameters can be obtained by minimizing variance of experimental data for current from theoretical values. 
Knowledge of these four parameters one can predict various roughness feature of
roughness profile like: mean roughness factor ($R^*$), root mean square (rms) width of roughness ($h \;{\rm or}\;\sqrt{m_0})$), 
rms gradient ($\sqrt{m_2}$) and inverse rms curvature ($r_c=1/\sqrt{m_4}$). 
The roughness factor is a function of mean square gradient $m_2$, i.e., $R^*\approx \sqrt{\pi m_2/2}$ for large roughness, which in turn is a strong function of ${\ell}$ 
and a weak function of ${\it L}$. 
Figure 2  compares our theory for current transient with rough gold deposit on wires.
Equation (\ref{scale-II}) is also obeyed by other experiments on surfaces like replica or gold masking of surfaces like fractured steel, dental surface and liquid-liquid interface. 
This theory predicts various roughness and marphological features and tabulated in Table 1 along with $t_i$.
\vskip0.2cm\hskip -0.2in
\begin{tabular}{l}
\hline 
{\bf\small Table 1 Predicted morphological parameters \& $t_i$}\\ 
\hline   
\begin{tabular}{l|llcccc|c}
Data & $D_H$$^1$  &  $D_H$$^3$ & $R^*_e$ & h ($\mu$m) & $\sqrt{m_2}$ & $r_c$ (nm) & $t_i$ (ms) \\
\hline 
Fig 2 a (1) & 2.4 & 2.42& 50  & 1.79 & 39.90 & 0.40 & 0.023\\
Fig 2 a (2) & 2.2 & 2.19 & 3 & 2.19 & 2.25 & 269.3 & 0.350\\
Fig 2 a (3) & 2.0 & 2.02 & 1.1 & 0.50 & 0.34 & 3344.8 & 0.743\\
Fig 2 b (1) & 2.52 & 2.52 & 100 & 4.87 & 79.97 & 0.35 & 0.143\\
Fig 2 b (2) & 2.24 & 2.33 & 20& 4.65 & 15.96 & 7.99 & 0.522\\
Fig 2 b (3) & 2.32 & 2.32 & 20 & 4.76 & 15.95 & 8.12 & 0.546\\
\hline
\end{tabular}\end{tabular}
\newline
\intextsep0.5in
{\footnotesize{$D_H$$^1$ and $D_H$$^3$ are calculated using Eq.(1) and (\ref{scale-II}), respectively. 
Roughness factor $R^*_e$ experimentally measured from voltammetric experiment\cite{Ocon-Arvia91} and 
is used to constraint value of $T$ parameter in our calculations, 
Predicted width of interface($h$), root mean square gradient ($\sqrt{m_2}$) and 
average radius of curvature($r_c$) and  inner crossover time($t_i$). 
Upper cutoff length ($L$) is kept constant in all calculations i.e., ${\it L} =3\;\mu m$.}}

The central result, equation~(\ref{scale-II}), constitutes an elegant and simple test that any data must pass
to be called limited scale self-affine fractals. This model gives a very good description of large quantity of data 
which have the scaling region as well as region which deviate from it.
It is important to note that the slope of scaling region does not purely depend
on fractal dimension alone but also  on lower cutoff length scale of roughness and strength of fractality too.
Another important observation is that the intermediate time expression is excellent for the current transient in 
intermediate and long time regimes (i.e., $t>t_i$).

This formulation opens new avenues towards an understanding of diffusion -limited reaction rates at 
realistic rough surfaces with limited scales of fractality and inverse problem of obtaining fractal parameters. 
We demonstrate that the fractal dimension is not in itself sufficient to understand this problem. 
This work unravels the connection between the anomalous intermediate 
power-law regime exponent and the geometric parameters of limited scales of fractality i.e. $D_H$, ${\ell}$, $T$.
I believe that this will provide insights into many other diffusion-limited processes on random interfaces
and the generality of this approach will provide logical extension to several complex situations, like various transient
techniques in electrochemistry, finite  reaction rates at interface, coupled homogeneous reactions with heterogeneous 
interfacial reaction etc. As more data becomes available with clear power spectral characterization of roughness,
quantitative analysis should provide an improved way to understand roughness of several surfaces.

\vskip1cm
{\footnotesize {\bf Acknowledgements:} {Author thanks T. R. Seshadri and S. K. Rangarajan for their useful suggestions and University of Delhi for financial support.}}

\end{document}